# Visually moving objects to an arbitrary distance by a simple shifting cloak

**Jianguo Guan,\* Wei Li, Zhigang Sun, and Wei Wang**

*State Key Lab of Advanced Technology for Materials Synthesis and Processing, Wuhan University of Technology, Wuhan 430070, China*
*\*guanjg@whut.edu.cn*

**Abstract:** A rectangular "shifting cloak", which visually shifts the cloaked object for a certain distance from the original place, is proposed. Comparing with the previously proposed similar cloaks, this rectangular shifting cloak has a much simpler structure, for it is constructed with only four blocks of homogeneous materials. The shifting distance can be arbitrarily tuned from zero to infinity in principle, thus the shifted image can be either in or out of the cloak. Various kinds of objects that have different sizes, shapes and constitutive parameters can be cloaked using the same device. Numerical simulations are performed to verify the properties of the shifting cloak.

## 1. Introduction

Recently, Pendry et al.[1] proposed a coordinate transformation method for the designing of the invisibility cloaks. Comparing with other design methods,[2-4] the coordinate transformation method (or transformation optics) is more powerful and flexible. More importantly, the design of the transformation optics based invisibility cloaks is irrelevant to the cloaked object.[1] So we can design a universal cloak which is applicable to various objects, and never need to learn the specific knowledge of the object being cloaked. Definitely, the complexity of designing a cloak has been greatly reduced by the transformation optics. Therefore, the transformation optics draws great attention of many scientists, and various forms of invisibility cloaks emerge in a short time.[5-10]

The aim of the invisibility cloak is to conceal an object from an observer by rendering the object transparent. Nevertheless, considering the implementation of the cloak, making things transparent is neither the only way, nor the most appropriate way to conceal an object in some cases, due to the shortcomings of the invisibility cloaks such as the requirement of highly anisotropic materials, continuously changing of the constitutive parameters with position and existence of singularities. Up to date, most of the efforts are dedicated to rendering the object transparent. Few alternative approaches have been proposed to achieve concealment as well. For example, by mimicking a conducting carpet, a ground-plane cloak[11] is able to hide objects in the background of a conducting sheet. The ground-plane cloak is relatively much easier to realize even in visible regime because only traditional materials are required.[12-14] Unfortunately, it can not be applied in air or vacuum. Another strategy is the "illusion optics"[15] which "cancels" the image of an object and then "restores" a stereoscopic false image, making the cloaked object looks like something else in the background of air or vacuum. By doing so, one can hide the cloaked object as well as release a false target to confuse the detector. However, for canceling and restoring images, an "anti-object" corresponding to the object being hidden and a compressed version of the "object" to be restored should be imbedded in the cloak. Imaginably, it would be hard to design and fabricate such a cloak for complex objects. Recently, a device which optically shifts and magnifies the image of the cloaked object is also proposed as a good alternative way to achieve concealment.[16] But the design and fabrication of that cloak should also consider the object being cloaked. In addition, the shifting of the image is always accompanied by the magnifying effect: the farther distance we shift, the bigger the "object" looks like. Sometimes, the highly magnified image may fail to cheat the observer due to its unreality. In our another work,[17] by introducing a "minifying layer", the scaling effect on the shifted image can be arbitrarily controlled. More interestingly, if the device is set to have no scaling effect, arbitrary object can be cloaked without redesigning the cloak. However, the device is a bit complicated because it is constructed with multilayered inhomogeneous materials. Another team also proposes a cloak that visually shifts the image of a cloaked object without scaling effect by a multi-steps coordinate transformation.[18] However, the shifted place of the image seems within the limit of the cloak region. In addition, the transformation and structure of the cloak are complicated. Considering the implementation of the cloaking device, the structure should be made as simple as possible. [19-22]

In this work, we propose a "shifting cloak" which visually shifts the image of an object from one place to another arbitrarily selected place. This shifting cloak visually shifts an object without magnifying effect, and thus making the image more realistic and deceptive to the observers. Moreover, designing and fabrication of the cloak are independent of the object to be cloaked. Comparing with previous similar works[17-18] whose constitutive parameters change continuously with the position, the cloak here is quit simple because it involves only four blocks of homogeneous materials. This feature can greatly improve the performance and decrease the complexity of fabrication when we implement the cloak.[22]

## 2. Principal

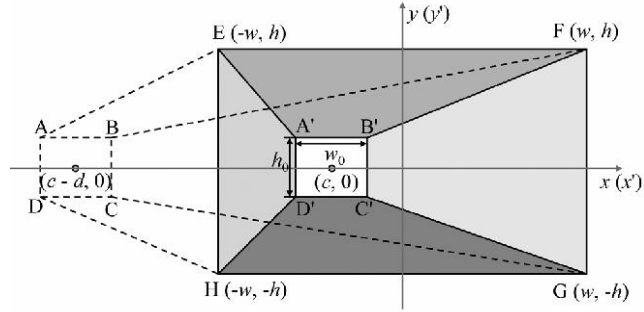

Fig. 1. Scheme of the coordinate transformation of a shifting cloak that visually shifts the cloaked objects to the left for a distance and out of the cloaking shell. Region ABCD, ABFE, BCGF, CDHG, ADHE in the virtual space are mapped onto the region A'B'C'D', A'B'FE, B'C'GF, C'D'HG, A'D'HE in the physical space, respectively.

For simplicity our analysis is restricted to the two dimensional (2D) case. As shown in Fig. 1, we want to create a rectangular cloak with dimension of $2w \times 2h$, containing a $w_0 \times h_0$ rectangular cloaked region. It translates the scattering of the cloaked object to the left for a distance of $d$. Two coincident Cartesian coordinate systems [$(x, y)$ and $(x', y')$] are defined for the virtual and physical space, respectively. Their origins are both at the geometrical center of the $2w \times 2h$ rectangle. The coordinate of the center of the rectangular cloaked region is $(c, 0)$. The coordinate transformation equations can then be expressed as:

for region A'D'HE (the left region):

$$x' = \frac{(w - w_0 + c)x + dw}{w - w_0 + c - d}, y' = y, z' = z \qquad (1)$$

for region A'B'FE (the top region):

$$x' = x + \frac{d(h - y)}{h - h_0}, y' = y, z' = z \qquad (2)$$

for region B'C'GF (the right region):

$$x' = \frac{(w - w_0 - c)x + dw}{w - w_0 - c + d}, y' = y, z' = z \qquad (3)$$

for region C'D'HG (the bottom region):

$$x' = x + \frac{d(h+y)}{h-h_0}, y' = y, z' = z \tag{4}$$

for region A'B'C'D' (the center region, i.e., the cloaked region):

$$x' = x + d, y' = y, z' = z. \tag{5}$$

By using the coordinate transformation method together with the coordinate transformation equations, it is feasible to deduce the material parameters of the cloak.[4] Taking the left part for example, its material parameters in the vacuum or air background can be given as:

$$\varepsilon_1^{'} = \mu_1^{'} = diag(K, \frac{1}{K}, \frac{1}{K}) \tag{6}$$

where $K = (w-w_0+c)/(w-w_0+c-d)$. It is worth noting that the material parameters expressed by Eq. (6) are spatially invariant, finite and non-zero. This means the material is homogeneous and non-singular. After deriving the material parameters of the rest parts, one can find that they are all spatially invariant and non-singular. In fact, during the transformation all the deformations proceed in one direction (the $x$ direction) homogeneously. As a result, the transformation equations are all linear. The cells of Jacobian matrix are all constant. Naturally, the cloak is constructed with only homogeneous materials. Interestingly, for the region A'B'C'D', i.e., the cloaked region, we have $\varepsilon_5^{'} = \mu_5^{'} = 1$ if the medium in region ABCD (in the virtual space) is vacuum or air. Because there is no deformation happened in that region during the transformation, the medium in region ABCD in the virtual space will transform into the physical space without any changes. That means everything in region A'B'C'D' of the physical space is visually equivalent to the same thing but at different place in the virtual space. Therefore, the shifting cloak is applicable for any object, no matter what the size, shape and constitutive materials are, and how it is placed in the cloaked region.

For controlling the shifting distance of the cloak, one just need let the parameter $d$ equal to the desired value. Then the material parameters of such a cloak are derived from Eqs. (1)-(4). In principle, $d$ can be an arbitrary positive real number from zero to infinity, corresponding to the shifting distance from zero to infinity. One can easily proves that in fact $d$ can also be negative real numbers, and in this case the cloaked object is visually shifted to right side for a distance of $|d|$. However, from Eq. (1) and (3) we know that $d$ can not be $w - w_0 \pm c$ for in these cases there will be singular.

### 3. Numerical simulation and analysis

With the above material descriptions, we can verify the properties of the shifting cloak by full wave numerical simulations. For simplicity, we consider the simulations in transverse electric (TE) modes. The parameters of the shifting cloak we consider are $w = 5$ m, $h = 3$ m, $a = 1$ m, $b = -1$ m and $d = 6$ m respectively. With these parameters, the shifting cloak would shift the scattering of the cloaked object to the left direction for a distance of 6 m. The wavelengths of the electromagnetic (EM) waves in all the simulations are 3 m in free space. Figure 2(a) shows the scattering pattern of the cloak when a perfect electric conducting (PEC) column with a square cross section is just fitted in the cloaked region and a plane wave is incident from left to right. For comparison, the same square PEC column without the shifting cloak is translated to the left direction for a distance of 6 m, and then the scattering pattern is presented as Fig. 2(b) under the same wave irradiation. According to the theoretical prediction, the above two cases will have the same far-field scattering pattern. Clearly, the two patterns in Fig. 2(a) and (b) are almost the same, indicating that the shifting cloak work properly as we predicted.

If we change the shifting distance $d$ to be a value of -3, as we have discussed in previous section, the equivalent virtual image will be at the right side and keeping a distance of 2 m with respect to the cloaked object. To verify the shifting effect of this cloak, numerical simulations have also performed. Figure 2(c) and (d) show the scattering patterns of the cloaked object and the equivalent object, respectively, when a plane wave is incident from top to bottom. It is clear that they are almost the same. Since the shifting distance is very short, the virtual image is still "in" the cloak region.

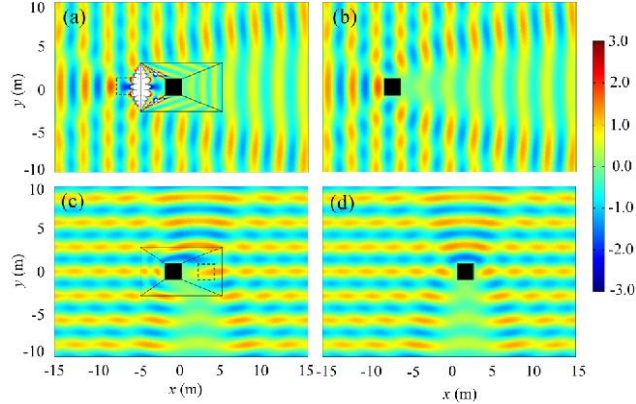

Fig. 2. Scattering pattern of (a) a square PEC column which is fitted into the cloak which is designed to visually shift an object to the left for a distance of 6 m, (b) remove the cloak in (a) and translate the square PEC column to the left for a distance of 6 m, when a TE plane wave is incident from left to right, (c) a square PEC column which is fitted into the cloak which is designed to visually shift an object to the right for a distance of 3 m, (d) remove the cloak in (a) and translate the square PEC column to the right for a distance of 3 m, when a TE plane wave is incident from top to bottom. For clarity, the square PEC columns are colored in black. The boundaries of the virtual images in (a, c) are outlined by dashed lines.

Some previously proposed cloaks[10,15-16] are related to the object being cloaked. If so, one must know the accurate shape and material properties of the object. The design and fabrication are thus greatly complicated. Fortunately, the shifting cloak here is applicable for objects with any shape and material properties as long as it can be putted into the cloaked region. To verify this property of the shifting cloak, we use the same cloak in Fig. 2(a) to shift the scattering of a rectangular dielectric column which is half size of the square PEC column in Fig. 2 and of parameters of $\mu = 1$ and $\varepsilon = 3$. The field distribution of the dielectric in the cloak when a plane wave is incident from left to right is shown in Fig. 3(a). According to the design of the cloak and the theoretical prediction, the cloaked object should scatters like it was shifted to the left for a distance of 6 m. Therefore we remove the cloak and shift the small dielectric to that position and get the scattering pattern which is presented as Fig. 3(b). The two coincident patterns in Fig. 3(a) and (b) confirm that the shifting cloak can be applied on that small dielectric.

From the above numerical simulations, we know that the rectangular shifting cloak is available to visually shift an object from its original place to anther place, regardless of the size, shape and constitutive of the object. When we design such a cloak, we can decide where the cloaked object is shifted to, either in or even out of the cloak. Besides, the structure of the shifting cloak is quit simple all along, because it is only composed of four blocks of homogeneous materials. And no extreme materials are involved such as materials with singularities. All related materials are obtainable with the assistance of the metamaterial technology. This shifting cloak can be used to hide the real position of an object and mislead the observer, especially when the virtual image of the cloaked object is shifted far from the object and out of the cloak. However, left-handed materials should be needed to achieve this unless the shifting distance is very short so that the virtual image is still in the cloak region.[18]

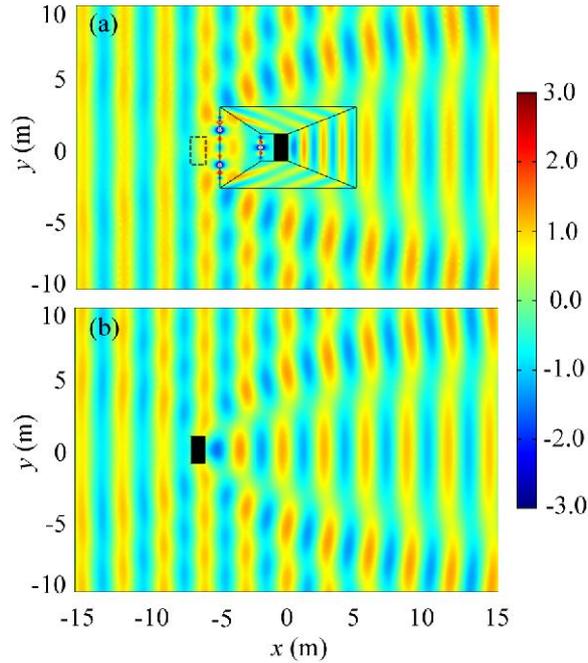

Fig. 3. Scattering pattern of (a) a rectangular dielectric column which is half size of the square PEC column and of parameters of $\mu = 1$ and $\varepsilon = 3$ and placed at the right side of the cloaked region, (b) remove the cloak in (a) and translate the dielectric to the left for a distance of 6 m, when a TE plane wave is incident from left to right. The cloak is the same as in Fig.2. For clarity, the rectangular dielectric column is colored in black. The boundary of the virtual image in (a) is outlined by dashed line.

## 4. Conclusion

In summery, we have derived a "shifting cloak" that shifts the scattering of the cloaked object, by using the coordinate transformation method. Numerical simulations have confirmed that various objects with different shapes and constitutive parameters can use this shifting cloak to hide themselves and simultaneously create false images look like the same with themselves at a distance to confuse the observers. It also shows that the shifted distance can be controlled easily by a parameter. The simple structure of the cloak makes it feasible to implement.

**Acknowledgements**

This work was supported by the National High-Technology Research and Development Program of China under Grant No. 2006AA03A209, the Young Teacher Grant from Fok Ying Tung Education Foundation under Grant No. 101049 and the Ministry of education of China under Grant No. PCSIRT0644.